\documentclass[]{aa}
\usepackage{graphicx}
\usepackage{txfonts}
\begin{document}
\title{Can the third dredge-up extinguish hot-bottom burning in massive 
AGB stars?}

   \author{Paola Marigo}
   \institute{Dipartimento di Astronomia, Universit\`a di Padova, 
		Vicolo dell'Osservatorio 3, 35122 Padova, Italia}
   

   \date{Received 20 November 2006 / Accepted 1 February 2007}

 \abstract{
Marigo (2002) has highlighted the crucial importance 
of molecular opacities in modelling the evolution of AGB stars at varying 
surface C/O ratio. In particular, it has been shown the large inadequacy of 
solar-scaled opacities when applied to models of carbon stars, and hence the 
need for correctly coupling the molecular opacities to the current surface 
chemical composition of AGB stars.
The aim of the present follow-up study is  
to investigate the effects of variable molecular opacities 
on the evolutionary properties of luminous
AGB stars with massive envelopes, i.e. with initial masses 
from $\approx 3.5 M_{\odot}$ up to
$5-8\, M_{\odot}$, 
which are predicted  to experience both the third dredge-up 
and hot-bottom burning.
It is found that if the dredge-up of carbon is efficient enough to
lead to an early transition from C/O $\,<1$ to C/O $\,>1$, then
hot-bottom burning may be weakened, extinguished, or even prevented.
The physical conditions for this occurrence are analysed and a few 
theoretical and observational implications are discussed.
Importantly, it is found that the inclusion of variable molecular 
opacities 
could significantly change the current predictions for the chemical
yields contributed by intermediate-mass AGB stars, with $M\simeq 3.5
- 4.0\, M_{\odot}$, that make as much as $\sim 30-50\,\%$ of all stars
expected to undergo hot-bottom burning.
}

\authorrunning{P. Marigo}   
\titlerunning{Extinction of Hot Bottom Burning by the Third Dredge-up}
\maketitle
%

\section{Introduction}
During the Thermally-Pulsating Asymptotic Giant Branch 
(TP-AGB) phase, intermediate-mass stars (with initial masses 
$3.5 M_{\odot} \la M \le M_{\rm up} = 5-8 M_{\odot}$, depending on 
metallicity and model details) experience a particularly rich nucleosynthesis,
whose products are exposed to the surface via 
two main processes, 
namely: the third dredge-up, and hot-bottom burning (hereinafter also HBB).

The former injects into the envelope elements like 
helium  and other species produced during 
thermal pulses, essentially  
primary carbon and s-process elements 
(Iben \& Renzini 1983; Busso et al. 1999).
The formation of carbon (C-) stars of N type (with surface C/O$>1$ and 
not belonging to binary systems) is naturally explained in this way.

The latter process corresponds to  nuclear H-burning that extends
from the radiative shell into the deepest and hottest
layers of the convective envelope during the quiescent 
inter-pulse periods. The main nuclear reactions at work 
are those of the CNO-cycle, 
so that the surface abundances 
of helium and nitrogen are expected to increase.
Heavier nuclei -- like some isotopes of neon, sodium, aluminum,
and magnesium -- can be produced from the Ne-Na and Mg-Al cycles, 
whereas some amount of fresh lithium can also be synthesised and 
brought to the surface via the so-called beryllium-transport mechanism 
(Cameron \& Fowler 1971), 
which requires suitable conditions,
involving the temperature at the base of the convective envelope,  
and  the time scales of nuclei and convective eddies
(see Lattanzio \& Wood 2003 for an overview of the HBB nucleosynthesis).

Actually, HBB has been invoked to explain (or to contribute explaining) 
a number of observational evidences, 
like for instance: 
the (almost) lack of visible luminous C-stars  
($M_{\rm bol} < -6$) and the concomitant deficiency  
of luminous M-stars  in the Magellanic Clouds (e.g. Costa \& Frogel 1996);
the existence among these stars of objects ($-6 > M_{\rm bol} > -7$)
with marked enhancement of surface lithium abundance
(e.g. Smith et al. 1995);
the overabundance of nitrogen and helium 
typical of type I Planetary Nebulae and some AGB stars 
(e.g. Peimbert \& Torres-Peimbert 1983; Smith \& Lambert 1990); 
the low value of the $^{12}$C$/^{13}$C ratio, 
close to the nuclear equilibrium value,
characterizing the group of luminous J-type carbon stars
(e.g. Lambert et al. 1986); the Na-O and Mg-Al anti-correlations characterising
the stars of galactic globular clusters in the framework of the primordial
scenario (Gratton et al. 2001);  the high isotopic ratios  
$^{26}$Al$/^{27}$Al displayed by some meteoritic oxide grains of pre-solar
origin (Mowlavi \& Meynet 2000). 
 
Several evolutionary models of massive AGB stars with HBB 
(say with $M \ga 3.5-4 M_{\odot}$, depending on metallicity)
have been calculated in recent years (e.g. Boothroyd \& Sackmann 1992; 
Bl\"ocker 1995; Forestini \& Charbonnel 1997; Marigo 1998; 
Frost et al. 1998;  Ventura et al. 1999; Karakas \& Lattanzio 2003;
Herwig 2004; Ventura \& D'Antona. 2005ab). 
Compared to earlier studies, 
these calculations mark a net progress in respect to many aspects, 
like for instance the adoption of larger nuclear networks including  many 
elemental and isotopic species, a more detailed description of convective 
turbulence and mixing, and
extensive calculations over wider ranges of initial stellar masses and
metallicities.

However, one standard choice common to all these models 
is the use of tables of molecular opacities 
-- for temperatures lower than $\sim 10\:000$ K  --  that are strictly
valid for solar-scaled abundances of elemental species heavier 
than helium (e.g. Alexander \& Ferguson 1994).
Then, for a fixed initial metallicity, 
interpolations are usually performed as a function of temperature, density, 
and hydrogen content. This implies that any change in the elemental 
and molecular concentrations (following the convective dredge-up and 
HBB) does not translate, as instead it should, into a change 
of the molecular opacities adopted to calculate  
the AGB model atmospheres.

This point of inconsistency has been clearly pointed out 
by Marigo (2002; see also Marigo et al. 2003b), 
who has constructed a routine to estimate the molecular opacities 
for whatever chemical composition of AGB envelopes. Typically,
the photospheric opacities are dominated by an opacity 
bump due to H$_2$O for  C/O$\,<1$, while as soon as C/O$\,>1$
this peak disappears and a prominent opacity 
bump due to the CN molecules develops.
 
The routine has been then incorporated into a static envelope model, which
is part of a code developed to calculate the synthetic TP-AGB evolution
(see Marigo et al. 1999 and references therein).   
By comparing the results of TP-AGB evolutionary calculations obtained 
with fixed  solar-scaled  to those with 
variable molecular opacities it turns out that
the surface C/O ratio plays a key-role in determining the evolutionary
properties of AGB stars. 

In Marigo (2002) most attention was devoted to study the impact of 
variable opacities on the evolutionary behaviour of low-mass stars 
(with masses up to $2.5 M_{\odot}$). It turns out that 
one of the major 
effects is the sudden and marked cooling of the 
evolutionary tracks in the H-R diagram as soon as C/O overcomes unity, 
that is at the transition from the oxygen-rich configuration (C/O$\, < 1$)
of M-type stars to that of  carbon-rich stars (C/O$\, > 1$). 
The attainment of lower effective temperatures
causes, in turn, other important  evolutionary consequences, namely:
earlier onset of the super-wind;
shorter duration of the TP-AGB and C-star phases, redder near-infrared colours,
reduced degree of carbon enrichment  hence lower chemical yield and C/O ratios
in carbon stars. Most of these predictions allows a better agreement 
with the observations of carbon stars.
Moreover, the relatively low 
C/O ratios ($\la 1$) exhibited by disk planetary nebulae can be more easily
explained with the aid of TP-AGB models including the third dredge-up 
and variable molecular opacities, as shown by Marigo et al. (2003a).

Following the introduction of
the new synthetic TP-AGB models into a population synthesis code 
Marigo et al. (2003b) have  shown
that the variable molecular opacities are the key to explain  
the prominent red tail of visible carbon stars seen in some near-infrared 
colour-magnitude (e.g. $K$ vs. $J-K$) diagrams, like those provided 
by the DENIS (Cioni et al. 2000) and 2MASS surveys (Nikolaev \& Weinberg 2000) 
towards the Magellanic Clouds.

The present paper pushes forward the analysis started with Marigo (2002)
and Marigo et al. (2003b),
by investigating the possible effects driven by variations of molecular  
opacities in the most massive AGB stars 
(with $M \ga 3.5  M_{\odot}$),  that experience both the third 
dredge-up and hot-bottom burning. 
In particular it will be shown that, if during the early stages 
of the TP-AGB phase the third dredge-up 
is able to produce C/O$>1$ in the envelope, then this 
may prevent, or weaken and eventually extinguish the nuclear rates 
associated with HBB. 

It follows that the third dredge-up and HBB produce reciprocal
interference.
In fact, on one side, it has been long predicted 
(e.g. Renzini \& Voli 1981; Boothroyd \& Sackmann 1992) that
HBB can delay or even prevent the formation of high-luminosity 
C-stars, by destroying the dredged-up carbon in favour of nitrogen.
On the other side, the present study shows, for the first time, 
that the third dredge-up itself may contrast the development of HBB.
This means that the two competing processes 
are intimately related one to each other, 
in a more complex manner  than considered so far.  

The structure of the paper is as follows.
Section~\ref{sec_envmod} details the envelope model adopted in the
present study. The sensitiveness of luminous and massive stellar
envelopes to increasing surface C/O ratio is first explored in 
Sect.~\ref{sec_statenv} with the aid of envelope integrations, and then
better investigated in Sects.~\ref{sec_evolcalc} and 
\ref{sec_depmz} where a few synthetic
TP-AGB calculations are presented for selected values
of the stellar initial mass and metallicity. Next,  
in Sect.~\ref{sec_lithium}, the analysis
is focused on the impact of
variable molecular opacities in changing  
the predicted evolutionary properties of a particular group of stars, i.e.
the lithium-rich carbon stars, while Sect.~\ref{sec_weight} weighs the
significance of the new results in the context of the theory of
stellar populations. The reader is made aware of a few cautionary
remarks on possible limits of the present work in Sect.~\ref{sec_caution}, 
whereas a closing  summary is provided in Sect.~\ref{sec_final}.

\section{Envelope integrations}
\label{sec_envmod}
The calculations presented in this paper are based on 
numerical integrations of a static envelope model, which is also a 
key-ingredient of the synthetic TP-AGB model already developed 
by Marigo et al. (1996, 1998, 1999).

Let us just briefly outline here the basic scheme.
Envelope integrations are carried out from the photosphere
down to mass-coordinate of the H-exhausted core.
The determination of the unknown functions 
$r$, $P_{r}$, $T_{r}$, $L_{r}$ across the envelope is obtained by solving
the stellar structure equations, once four boundary conditions
are specified.  
Two conditions    
naturally derive from the integration of the photospheric 
equations for $T$ and $P$ down to the bottom of the photosphere
(Kippenhahn et al. 1967).
The remaining two conditions involve the radius and
the local energy flux at the mass-coordinate of the He-core
(see  Marigo et al. (1998), Marigo (1998) for all the details).

Given the mass of the core, 
and the mass and chemical composition of the envelope, 
this procedure allows to single out the main envelope parameters, namely
the luminosity $L$ and the effective temperature $T_{\rm eff}$.

It is important to remark that our model accounts for 
nuclear energy sources possibly operating in the deepest envelope
layers. The common prescription of envelope models $L_{r} = constant = L$ 
is therefore replaced with the equation of energy balance 
$\partial L_{r}/{\partial M_r}  =  \epsilon_r$ (with usual meaning
of the quantities).
In this way we can consistently follow 
hot-bottom burning in the more massive AGB stars, 
with respect to both the CNO-cycle nucleosynthesis 
and the related over-luminosity
effect that makes these models more luminous than predicted by the
core mass-luminosity relation for the same $M_{\rm c}$
(see  Marigo et al. 1998, Marigo 1998 for a detailed discussion).

The basic physical inputs employed in our envelope model are 
as follows:
Nuclear reaction rates for the p-p chains  and 
CNO-bicycle are taken from  the compilation 
by Caughlan \& Fowler (1988), the screening factors are given
by Graboske et al. (1973). The adopted mixing-length parameter is
$\alpha = 1.68$. 

As far as the gas opacities are concerned we use those 
calculated by Iglesias \& Rogers (1996) (OPAL) at high temperatures 
($T > 10^{4}$ K). 
At lower temperatures ($T< 10^{4}$ K) we adopt the opacity
routine developed by Marigo (2002) choosing between two
alternatives for the adopted chemical composition:
\begin{itemize}
\item 
 a solar-scaled mixture of elemental species at any metallicity,
as in  Alexander \& Ferguson (1994) opacity tables
(hereinafter also $\kappa_{\rm fix}$);
\item 
the current mixture of chemical elements as it results from 
the nucleosynthesis and mixing events occurring 
during the TP-AGB evolution of the model under consideration
(hereinafter also $\kappa_{\rm var}$).
\end{itemize}

\begin{figure}
\resizebox{\hsize}{!}{\includegraphics{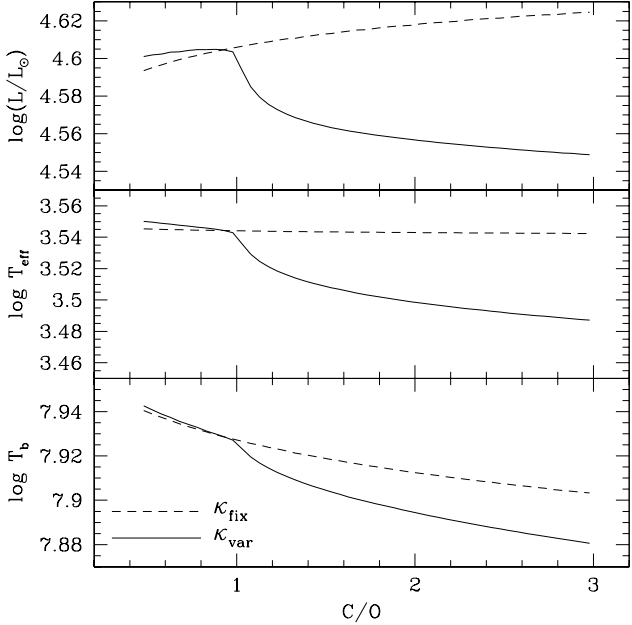}}
\caption{From top to bottom: 
predicted behaviour 
of the temperature at the base of the convective envelope, 
effective temperature, and luminosity 
at varying surface C/O ratio. Carbon abundance (by number) 
is made to increase in steps of 0.05, while that of oxygen 
is kept constant.  
The curves are the results of envelope integrations, 
assuming a total stellar mass of 4.5 $M_{\odot}$, 
a core mass of 0.98 $M_{\odot}$, an initial metallicity $Z=0.004,
$, and adopting either 
$\kappa_{\rm fix}$ or $\kappa_{\rm var}$.
See text for more explanation }
\label{fig_env}
\end{figure}
 
\section{Sensitiveness of extended AGB envelopes to the C/O ratio} 
\label{sec_statenv}

We will now investigate the effects 
produced by an efficient dredge-up of carbon
on the envelope structure of massive and luminous AGB models, that are 
also predicted to experience HBB. 

To this aim, Fig.~\ref{fig_env} shows the results 
of some test calculations, carried out by means of  
envelope integrations with either $\kappa_{\rm fix}$
or $\kappa_{\rm var}$ (see Sect.~\ref{sec_envmod}). We assume
a core mass $M_{\rm c} = 0.98\, M_{\odot}$, a total stellar mass
$M = 4.5\, M_{\odot}$, and
an initial chemical composition of the envelope 
[$Z_{\rm i}=0.004, Y_{\rm i}=0.24$], 
with $Z_{\rm i}$, $Y_{\rm i}$, $X_{\rm i}=1-Y_{\rm i}-Z_{\rm i}$ 
being the abundances (in mass fraction) of 
metals, helium and hydrogen, respectively. 

Initially, the elemental abundances 
of all species heavier than helium are assumed
to be solar-scaled according to compilation of solar abundances 
by Grevesse \& Noels (1993). 
It follows that the starting C/O ratio in the envelope is $\sim 0.48$.
Then, envelope integrations are performed at varying C/O ratio, which
is made increase from below to above unity by adding carbon ($^{12}$C) while
keeping oxygen fixed to its solar-scaled abundance.
We specify that in all models discussed in the paper the C/O ratio 
is calculated 

\begin{equation}
\nonumber
\frac{\rm C}{\rm O} = \frac{X(^{12}\rm{C})/12+X(^{13}\rm{C})/13}
{X(^{16}\rm{O})/16+X(^{17}\rm{O})/17+X(^{18}\rm{O})/18}
\end{equation}

where the $X$s in the right-side member 
denote the elemental abundances of the C and O isotopes (mass fractions).
The excess of carbon 
over its solar-scaled
value, $\Delta C = X(^{12}$C$)- X(^{12}$C$)_{\odot}$,  
is compensated by reducing the hydrogen abundance by the same amount, 
that is $X = X_{\rm i} - \Delta C$, which also implies that $Y=Y_{\rm i}$.

Figure \ref{fig_env} displays the behaviour of relevant quantities -- 
i.e. temperature at the base of the convective envelope $T_{\rm bce}$, 
effective temperature $T_{\rm eff}$, and stellar luminosity $L$ -- 
as a function of C/O.

The first point to be noticed is 
that, while in the $\kappa_{\rm fix}$ case the  
variables show a smooth and regular trend over the whole C/O range,  
a marked slope change shows up 
at the transition from C/O $<1$ to C/O $> 1$ in the $\kappa_{\rm var}$ 
curves.
Second, with the $\kappa_{\rm fix}$ assumption the variations of 
$T_{\rm bce}$, $T_{\rm eff}$, and $L$ are globally smaller, 
and they may even proceed
to the opposite direction compared to the 
$\kappa_{\rm var}$ case, as is for the the surface luminosity.
These differences can be explained as follows.

On one hand, the steady brightening of the model with $\kappa_{\rm fix}$
is essentially due to a cycling effect, that is 
more and more carbon is progressively 
made available to enter the CN cycle at the base 
of the convective envelope, thus increasing the efficiency of nuclear 
reactions.
Despite the modest decrease of  
$T_{\rm bce}$, which would tend to lower the nuclear rates, 
this carbon cycling effect dominates in determining the behaviour of L.
In the outermost atmospheric layers 
the effect of the increased carbon abundance 
on the temperature stratification becomes almost negligible so that
$T_{\rm eff}$ remains practically constant.

On the other hand, the inflection points at C/O$\,\sim 1$, that are
present in all the three curves with $\kappa_{\rm var}$, are caused by 
the sudden 
change of the dominant opacity sources at low temperatures ($T < 6000$ K),
reflecting, in turn, the abrupt alteration of the molecular concentrations
(see Marigo 2002).
In fact, the development of the large CN opacity bump 
for C/O$\,>1$ causes a marked decrease of both $T_{\rm eff}$ and $T_{\rm bce}$.
In this case, the cooling effect of the envelope base prevails over 
the injection of new carbon into the CN cycle, so that the efficiency
of nuclear energy generation ($\epsilon_{\rm CNO} \propto T^{15-20}$) 
becomes lower with consequent decrease of 
the surface luminosity.

\begin{figure*}
\resizebox{\hsize}{!}{\includegraphics{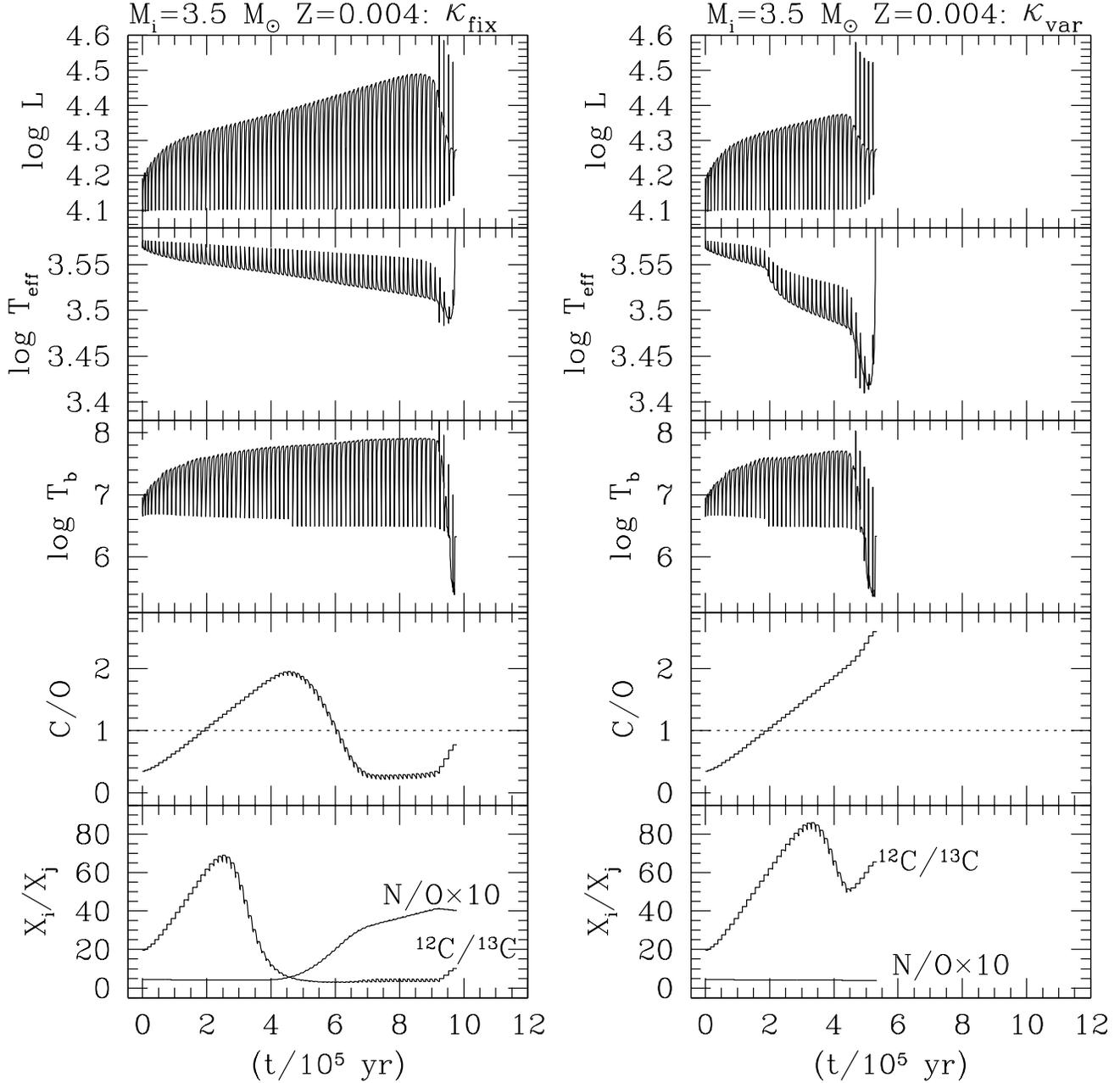}}
\caption{
From top to bottom: temporal evolution of luminosity,
effective temperature, temperature at the base of the convective
envelope, surface C/O, $^{12}$C$/^{13}$C, and N/O ratios during the TP-AGB 
phase of a $(M_{\rm i}=3.5,\,Z=0.004)$.
Calculations are carried out from the first thermal pulse
up to the  ejection of the whole envelope according to
Vassiliadis \& Wood (1993) formalism for mass loss}
\label{fig_35z004}
\end{figure*}

In summary, from these simple test calculations we already derive 
the indication that molecular opacities -- coupled to the actual chemical
 abundances in the envelope -- could significantly influence the structure 
and nucleosynthesis of massive AGB stars by affecting 
the efficiency of HBB. 
Anyhow, it should be noticed that 
these preliminary results indicate how the structure of a
stationary envelope reacts to changes in its chemical abundances,
hence in molecular opacities,
while all other parameters (e.g. core mass, total stellar mass) 
are kept fixed.  
The next necessary step is to analyse the impact on the evolutionary 
properties of massive AGB stars, which is done in the next section. 
   
\section{TP-AGB evolutionary calculations}
\label{sec_evolcalc}
We have calculated the TP-AGB evolution of  a few 
stellar models experiencing both the third dredge-up and HBB. 
Calculations are performed  with the aid of the  
synthetic TP-AGB model as recently revised 
by Marigo \& Girardi (2006, and references therein), to whom we refer for
all details (see also Sect.~\ref{sec_envmod}).

Suffice it here to recall that
the initial conditions at the first thermal pulse -- 
i.e. total mass, core mass, 
luminosity, surface chemical composition -- are taken from 
Girardi et al. (2000).
The third dredge-up is parametrized in terms of three quantities, namely:
the minimum core mass  
required for the occurrence of convective dredge-up, 
$M_{\rm c}^{\rm min}$;  the dredge-up efficiency, $\lambda$; 
and the chemical composition  of the 
dredged-up material.

In practice, following the indications from full AGB calculations 
(Karakas et al. 2002), 
in all models here considered the third dredge-up is assumed
i) to occur since the first thermal pulse,  and ii) to be quite efficient,
typically with $\lambda \approx 0.9-1.0$. 
The adopted inter-shell abundances depend on the core mass growth 
 as described in 
Marigo \& Girardi (2006), reaching standard maximum values  
[$X($He$)\simeq 0.73$, $X(^{12}$C$)\simeq 0.25$, $X(^{16}$O$)\simeq 0.02$].

As already mentioned in Sect.~\ref{sec_envmod}, HBB
is followed in detail with the aid of a complete
envelope model, including the most important reactions rates of the 
CNO bi-cycle. The over-luminosity effect due to 
HBB and the consequent break-down of the core mass-luminosity
relation are taken into account. 
Mass loss is described with the formalism
by Vassiliadis \& Wood (1993), under the assumption that variable AGB stars
are fundamental-mode pulsators and applying the delay in the onset of 
the super-wind for $M \ga 2.5\, M_{\odot}$ (their equation~5).

Let us first consider the case of a (3.5 $M_{\odot}, Z=0.004$) model.
Figure~\ref{fig_35z004} displays the evolution of 
characteristic quantities -- 
i.e. luminosity, effective temperature, temperature
at the base of the convective envelope, and a few surface elemental ratios --
from the first thermal pulse up
to the termination of the TP-AGB phase, marked by 
the complete ejection of the envelope.
The results are shown for both opacity prescriptions, 
$\kappa_{\rm fix}$ (left panel)   and $\kappa_{\rm var}$  (right panel).

From the comparison between the two models we notice that  significant 
differences show up  as soon as the surface 
C/O ratio overcomes unity as a consequence of the third dredge-up.

First of all, we recover the same photospheric cooling effect
found for lower mass AGB models (say, with $M < 3.5  M_{\odot}$) 
that undergo the transition 
to the C-star domain, as fully discussed  by
Marigo (2002) and Marigo et al. (2003).
In fact,  
a sudden jump to lower $T_{\rm eff}$ is seen to occur
in the  $\kappa_{\rm var}$ model at C/O$\ga 1$, 
while a smooth temporal behaviour  of  $T_{\rm eff}$ still characterises 
the evolution with $\kappa_{\rm fix}$.
The attainment of  lower $T_{\rm eff}$ by the  $\kappa_{\rm var}$ model 
yields longer pulsation periods,
hence favouring an earlier onset of the super-wind regime of 
mass loss. As a consequence, the duration of 
the TP-AGB phase turns out to be reduced by almost a factor of 2.

In addition to this, the introduction of variable molecular opacities
is expected to impact significantly also on the nucleosynthesis
associated with HBB. This is the combined result 
of enhanced mass-loss rates and somewhat lower temperatures 
at the base of the convective envelope (see Sect.~\ref{sec_statenv}). 
Both factors concur
to weaken the efficiency of nuclear reactions.

As for the evolution of the C/O ratio, the following aspects should be 
noticed.
As we see in Fig.~\ref{fig_35z004} the  $\kappa_{\rm fix}$ model
first becomes  carbon-rich (C/O$\,>1$) because of the third dredge-up, 
and later it is converted back as 
oxygen-rich (C/O$\,<1$) by the CN-cycle operating 
in the inner layers of the convective envelope. This configuration
is held up to the end of the AGB phase.
On the contrary, in the  $\kappa_{\rm var}$ case, once the surface 
carbon abundance overcomes that of oxygen, 
the C/O ratio keeps on increasing so that at the termination of the 
TP-AGB phase the model is still carbon-rich.

The fact that in the $\kappa_{\rm var}$ model HBB is effectively 
extinguished soon after the transition to the C-star domain is clearly
indicated by the behaviour of other surface abundances, which are displayed
in the bottom panels of Fig.~\ref{fig_35z004}.
The $^{12}$C$/^{13}$C ratio, for instance, 
is prevented from attaining nuclear equilibrium
($\simeq 3-4$), which is instead reached and maintained for long 
by the $\kappa_{\rm fix}$ model. More remarkably, the N/O ratio remains 
practically constant over the entire TP-AGB evolution with $\kappa_{\rm var}$,
implying that the CN cycle has never reached the necessary  
thermodynamical conditions (i.e. temperature and density) 
to convert carbon into nitrogen. Some nitrogen production is, instead, 
predicted for the model with  $\kappa_{\rm fix}$, as shown by the rising part 
of the N/O curve.
 
In summary, we expect that the efficiency of HBB may be reduced 
even significantly if,
during the early stages  of its TP-AGB evolution, 
a massive  AGB star experiences efficient carbon dredge-up 
such as to  become a carbon star. 
Then, the enhanced molecular opacities, associated with a carbon-rich
chemical composition, importantly 
affect the temperature stratification of the layers
above the core. A cooling effect takes place both at base of the convective
envelope, i.e. lowering  $T_{\rm bce}$, and at the atmosphere, 
i.e. lowering $T_{\rm eff}$.
The way the two factors concur to weaken, and possibly to extinguish HBB, 
is different.
The decrease of $T_{\rm bce}$  has, clearly, a direct effect on
the nuclear reaction rates of HBB, while the 
decrease of $T_{\rm eff}$ affects the HBB efficiency by altering 
the evolutionary path of the TP-AGB model.
In fact, a lower effective temperature favours  the attainment of larger
mass-loss rates, hence an earlier extinction of HBB due to the reduction
of the envelope mass by stellar winds.

It is clear that the prevailing effect between 
the two depends on several parameters such as:
mass loss  and its dependence on $T_{\rm eff}$, 
dredge-up efficiency, stellar initial mass and metallicity.
In Sect.~\ref{sec_depmz} we will discuss exemplifying  
results for various choices of the stellar mass
and metallicity.

\section{Dependence on mass and metallicity}
\label{sec_depmz}

\begin{figure*}
\resizebox{\hsize}{!}{\includegraphics{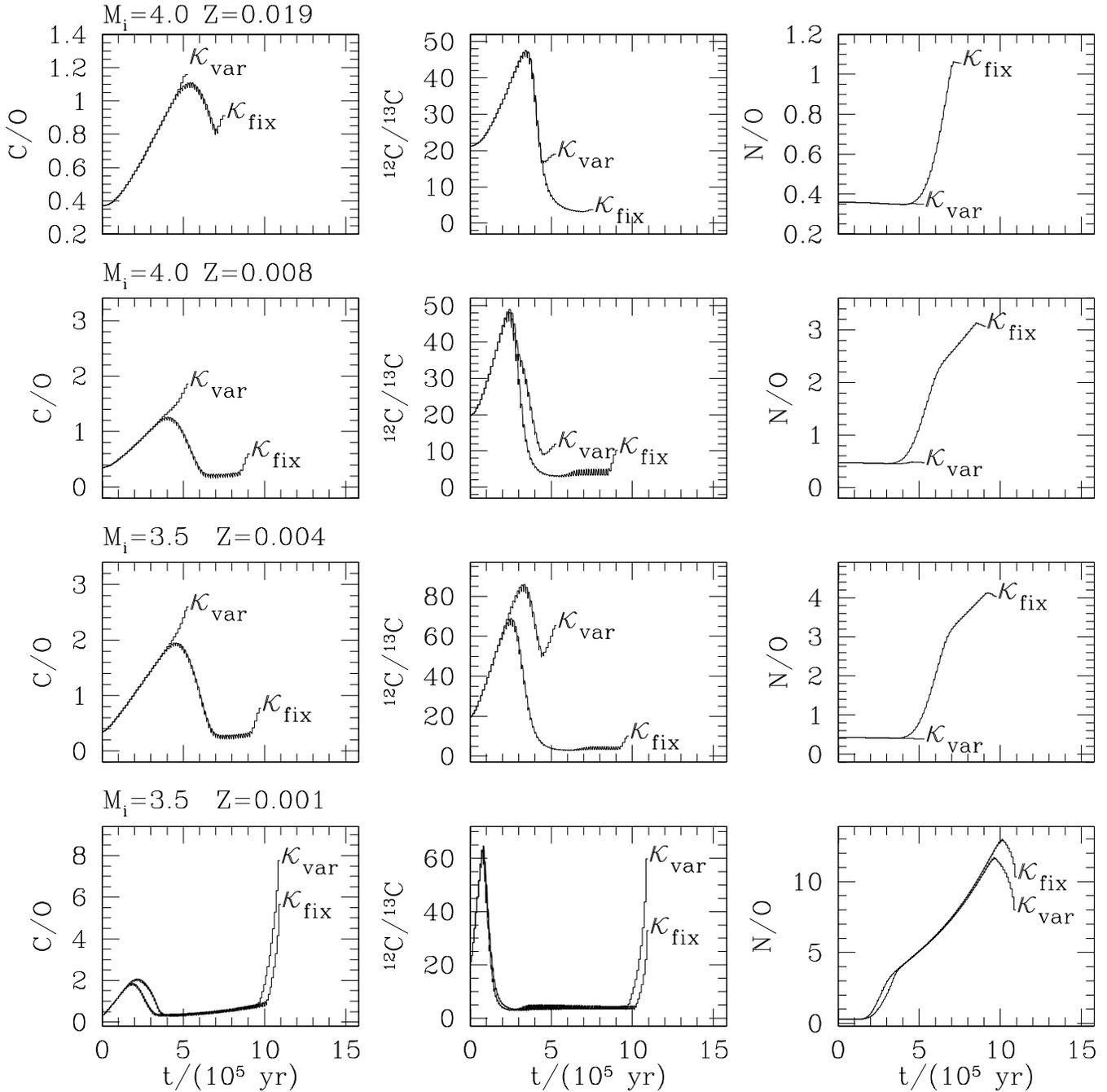}}
\caption{Predicted temporal evolution of surface elemental
ratios (by number) during the whole TP-AGB evolution 
of a few selected models with varying initial stellar
mass and metallicity, as indicated.
Results are displayed for both choices of molecular
opacities, $\kappa_{\rm fix}$ and $\kappa_{\rm var}$.
Both the third dredge-up and HBB are at work in these models
 }
\label{fig_xabund}
\end{figure*}

The interference of the third dredge-up against the efficiency of HBB 
develops to a different extent  depending on several factors, such as
initial stellar mass, dredge-up characteristics 
(efficiency $\lambda$ and chemical composition of the inter-shell), 
mass loss prescription, and metallicity.

In this study we limit to explore the sensitiveness of the results
on initial stellar mass and metallicity.
A few cases are presented in 
Fig.~\ref{fig_xabund},  that shows  the time evolution of the surface
chemical ratios C/O,  $^{12}$C$/^{13}$C, and N/O, adopting  both 
$\kappa_{\rm fix}$ and $\kappa_{\rm var}$ prescriptions.

The three ratios give information on the strength of HBB and its competition
with the third dredge-up, the behaviour of C/O showing directly if a star
should appear either as an M-type or as a C-type giant,  where the trends 
of $^{12}$C$/^{13}$C  and  N/O
measure the efficiency of HBB to activate the first 
nuclear reactions of the CN cycle (via the conversion  
of $^{12}$C to $^{13}$C), and subsequently to produce  
primary N at the expense
of the newly dredged-up C.  

The choice of the initial masses is determined by the requirement that,
in the  $\kappa_{\rm fix}$ case, the models actually make an early
transition to the C-star domain, which is the necessary condition 
to investigate a possible opacity effect in the corresponding  
$\kappa_{\rm var}$ calculations.
In practice, this means to consider models experiencing 
a moderate HBB, that is with initial masses in the lowest part 
of the typical mass interval for its occurrence, i.e.
$M_{\rm i} \approx 3.5 M_{\odot}-4.0 M_{\odot}$, depending on metallicity.

From the inspection of Fig.~\ref{fig_xabund}
it turns out that the discussion made in Sect.~\ref{sec_evolcalc}
in respect to the  $(M_{\rm i}=3.5,\,Z=0.004)$ model applies
also to  the $(M_{\rm i}=4.0,\,Z=0.019)$ and $(M_{\rm  i}=4.0,\,Z=0.008)$ 
models: The adoption of variable molecular opacities leads to an effective
extinction of HBB shortly after the attainment of C/O$\,>1$.

Different is the case of the $(M_{\rm i}=3.5,\,Z=0.001)$ model (bottom panel),
 with the lowest metallicity here considered. 
The differences of the results  between 
$\kappa_{\rm fix}$ and   $\kappa_{\rm var}$ are just minor and both models
experience efficient HBB despite the fact that they become carbon-rich  during
the initial stages of their TP-AGB evolution.
This can be understood by considering that at decreasing metallicity
molecular concentrations get lower and lower because of i) the 
lower abundances of the involved atoms, and ii) the strong temperature
dependence of molecular formation on temperature.
In extremely metal-poor atmospheres the temperatures
may be so high that the molecules cannot be even assembled. 

\section{The evolution of lithium-rich carbon stars}
\label{sec_lithium}
%
\begin{figure}
\resizebox{\hsize}{!}{\includegraphics{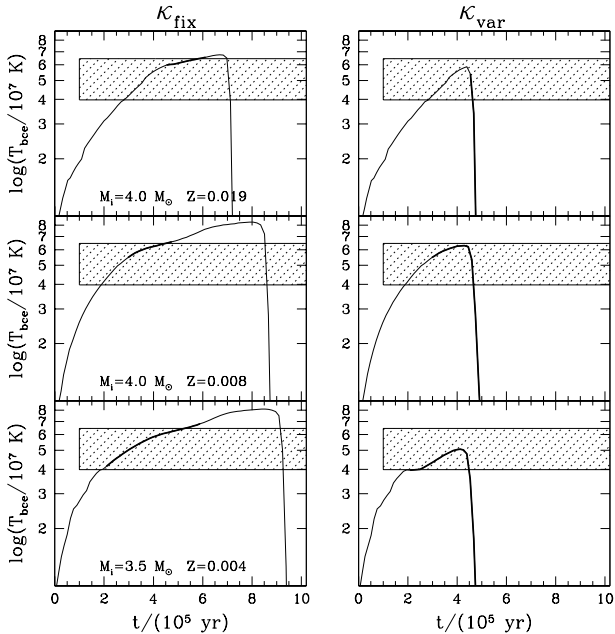}}
\caption{Temperature, $T_{\rm bce}$, at the base of the convective envelope
as a function of time during the TP-AGB evolution 
of stellar models with initial mass and metallicity as indicated.
For the sake of simplicity only the quiescent pre-flash stages are
considered in the construction of the plot. 
Results are shown for both $\kappa_{\rm fix}$ (left panels) and  
$\kappa_{\rm var}$ (right panels)
prescriptions. Thick lines correspond to C/O$\,>1$ at the surface.
The hatched strip is defined by the temperature conditions required to produce
lithium-rich carbon stars according to Ventura et al. (1999). 
See text for more explanation}
\label{fig_lith}
\end{figure}
As shown in the previous sections, the opacity effect driven 
by the third dredge-up is expected to be important 
for AGB stars that both are carbon rich and experience HBB. 
Interestingly, there is one group of observed objects 
that could meet these conditions, namely the luminous 
lithium-rich carbon stars.
In fact, among the most luminous carbon stars detected in the Magellanic 
Clouds  a few possess 
a significant enhancement of lithium abundance in the
magnitude range  $-5.5 \ga M_{\rm bol} \ga -6.5$ 
(e.g. Smith \& Lambert 1989, 1990). 

These stars are currently explained 
considering the combined effect of the 
the third dredge-up for the surface enrichment of carbon,
and the Cameron \& Fowler (1971) mechanism, operating under HBB conditions,  
for the synthesis and temporary survival of lithium at the photosphere.
Ventura et al. (1999)  address this question with the aid of full
AGB evolutionary models (with initial masses 3.3$\le M/M_{\odot} \le 4.0$). 
They show that  the chemical peculiarities of the high-luminosity lithium-rich 
C-stars can be reproduced only under strict structural conditions for the 
occurrence of HBB. 
Specifically, the temperature at 
the base of the convective envelope should vary within a narrow range, that is
$4 \times 10^7 \la T_{\rm bce} \la 6.5 \times 10^7$.
The lower limit is the minimum temperature required to activate the 
Cameron \& Fowler   mechanism for lithium production, the upper limit 
corresponds to the maximum temperature to preserve carbon against an efficient
destruction by the CNO-cycle. 
This temperature requirement would be fulfilled  
by AGB stars with  initial masses and luminosities within 
small ranges. Just in virtue of the predicted narrow luminosity domain 
these stars might be also used as independent distance-indicators, as
suggested  by Ventura et al. (1999).

In the light of the analysis just developed  
we can already envisage that the  
introduction of variable molecular opacities (not included in  
Ventura et al. (1999) calculations) may alter some of 
those findings.
It is not among the scopes of the present paper to perform a 
detailed analysis of all these issues, however 
our test calculations  show that the 
impact of variable molecular opacities 
may be important for models of  lithium- and carbon-rich
stars and future detailed models should include this physics. 

Figure~\ref{fig_lith} helps to illustrate this point. 
The evolution 
of $T_{\rm bce}$ is shown as a function of time 
during the TP-AGB phase of the same intermediate-mass models already 
discussed in the previous sections, assuming
either  $\kappa_{\rm fix}$ or  $\kappa_{\rm var}$.
 Following Ventura et al. (1999) the strict temperature conditions for
the formation of 
lithium-rich carbon stars define the hatched rectangular region.

A systematic difference affects the  results obtained 
with variable molecular opacities. While in models
with $\kappa_{\rm fix}$ the C-rich {\em and} Li-rich phase 
is characterised by a steady increase of
 $T_{\rm bce}$ which keeps on for some time even after the subsequent 
conversion to an oxygen-rich configuration, in models  
with $\kappa_{\rm var}$ the same phase always
intercept the maximum of $T_{\rm bce}$, after which an abrupt drop
is expected as a consequence of the onset of the super-wind.

As a consequence, we  expect significant changes in 
the predicted properties of these
stars,  such as the duration, the luminosity range, and the maximum 
 Li production of this chemically-peculiar  phase. 
Important differences should result also in respect to
the chemical yields from these stars.
In fact the  latter group of models with  $\kappa_{\rm var}$ 
is expected to enrich the ISM with newly synthesised carbon and lithium, 
while the former models with   $\kappa_{\rm fix}$ should
eject primary nitrogen and lithium. A quantitative prediction 
of the yields from these stars cannot be given at this stage 
and it is postponed to a future work.  

\section{Weighing the impact of the new results}
\label{sec_weight}
As already discuss in Sect.~\ref{sec_depmz}, 
the introduction of variable molecular opacities may importantly
affect AGB models experiencing both the third dredge-up and HBB,
within a relatively  narrow range of initial
stellar masses, roughly from 3.0  up to 4.0 $M_{\odot}$, depending on
metallicity and model assumptions.

To assess the significance of the new results 
we will exploit  
a few basic relations from the theory of stellar populations 
(e.g. Tinsley 1980).  
In a composite stellar population like a galaxy,
the expected number of TP-AGB stars with initial masses 
in the interval $[m, m+dm]$ is
expressed by
\begin{equation}
N(m)\, dm \propto \phi(m)\, 
\psi[T_{\rm G}-t(m)]\, t_{\rm TP-AGB}(m)\, dm
\label{eq_nm}
\end{equation}  
where $\phi(m)$ is the initial mass function (IMF), 
$\psi[T_{\rm G}-t(m)]$ is the star formation rate (SFR) evaluated at
the time of stellar birth  ($T_{\rm G}$ is the galaxy age  and $t(m)$
is the present stellar age), and $t_{\rm TP-AGB}(m)$ 
is the TP-AGB lifetime of a star with initial mass $m$. 

Let us now consider the typical mass range for the occurrence of HBB, 
say from $m=M_{\rm min}^{\rm HBB}$ to $m=M_{\rm up}$. 
The lower limit corresponds to the 
minimum stellar mass necessary to get $T_{\rm bce}$
high enough to activate nuclear H-burning in the deepest layers of the 
convective envelope, i.e. $\approx 1 \times 10^7$ K for the p-p chains or
$\approx 2 \times 10^7$ K for the CNO cycle.
Present stellar models (Romano et al. 2001)  
indicate that the production of lithium from AGB stars via the  
beryllium-transport mechanism
requires $M_{\rm min}^{\rm HBB} \approx 3.0 - 3.5 M_{\odot}$ as the 
metallicity increases in the 
range $Z=0.001-0.02$.

The upper 
limit $M_{\rm up}$ is the
maximum initial mass for a star to develop an electron-degenerate C-O
core, and its predicted value is sensitive to metallicity and 
model details, e.g. the treatment of convective
boundaries. In most cases $M_{\rm up}$ is expected to lie within 5 and 
8 $M_{\odot}$.

Relative to the mass range $[M_{\rm min}^{\rm HBB}, M_{\rm up}]$, 
{\em what is 
the expected fraction of AGB stars for which the present study could 
significantly alter the existing model predictions, i.e. 
in terms of evolutionary properties and chemical yields?}

To answer this question it is useful to calculate 
the cumulative  distribution function
\begin{equation}
F(M)=P(m\le M)=\frac{\int_{M_{\rm min}^{\rm HBB}}^{M} N(m)\, dm}
{\int_{M_{\rm min}^{\rm HBB}}^{M_{\rm up}}N(m)\, dm}
\end{equation}
where $N(m)\, dm$ is given by Eq.~(\ref{eq_nm}).
This latter gives the probability that an AGB star has a progenitor
with an initial mass
$m\le M$, with $m$ varying within the range  
$[M_{\rm min}^{\rm HBB}, M_{\rm up}]$.

\begin{figure}
\resizebox{\hsize}{!}{\includegraphics{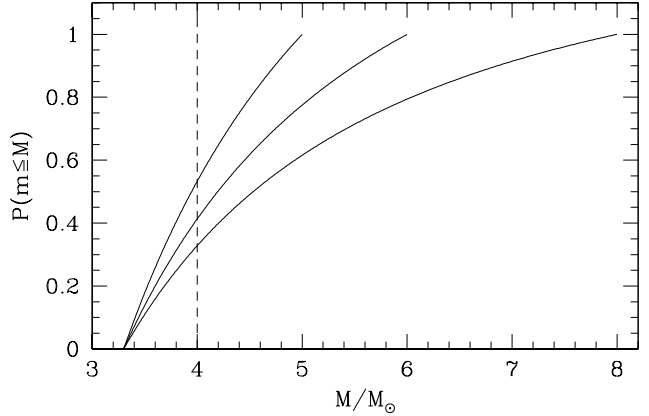}}
\caption{Cumulative distribution function defined by
  Eq.~(\ref{eq_nm}), yielding the probability to have an AGB 
star with initial mass $3.3\, M_{\odot} \le m \le M$. The distribution
is normalised to the total number of AGB stars with masses in the
  range $[M_{\rm min}^{\rm HBB}, M_{\rm up}]$, where  $M_{\rm up}$ is 
set equal to 5, 6, and 8 $M_{\odot}$ in the three plotted curves }
\label{fig_fracHBB}
\end{figure}
 
For the sake of simplicity, a first estimate is obtained 
assuming constant metallicity, constant SFR, 
 $t_{\rm TP-AGB}$ independent of the stellar mass, and adopting the Salpeter 
IMF ($\phi(m) \propto m^{-2.35}$). 
The results are shown in Fig.~\ref{fig_fracHBB} for three choices of
$M_{\rm up}$ and setting  $M_{\rm min}^{\rm HBB}=3.3\,  M_{\odot}$.
We see that in all cases $F(3.3 M_{\odot})=0$ and $F(M_{\rm up})=1$ 
by construction.

It turns out that  $F(4)=0.33, 0.42, 0.53$ 
for $M_{\rm up}=8,6,5\, M_{\odot}$ 
respectively, meaning that AGB stars with
initial masses in the interval $[3.3,4.0]\,M_{\odot}$ correspond to a
sizable fraction of whole massive range.
We remark that these are lower limits to the true values
since larger fractions are expected when taking the  
mass-dependence of $t_{\rm TP-AGB}$ into account, 
as the TP-AGB lifetime should decrease with
increasing stellar mass (e.g. Vassiliadis \& Wood 1993).

In summary, these simple calculations indicate that the impact
of the variable molecular opacities should apply to an important 
fraction of more massive AGB models.
For our particular class of models, with  $M_{\rm up}=5\, M_{\odot}$, 
more than half of the stars experiencing HBB should be affected by 
the results of the present study.

\section{Cautionary remarks}
\label{sec_caution}
The results of this study depend on the
assumed law for mass loss and its sensitiveness to changes in $T_{\rm eff}$.
In our exploratory calculations we have adopted the formalism by 
Vassiliadis \& Wood (1993) according to which, before the
onset of the super-wind, the mass-loss rate 
scales exponentially with the pulsation period 
and $P  \propto T_{\rm eff}^{-3.88}$.
This dependence on $T_{\rm eff}$  derives from the underlying 
period-mass-radius for the fundamental mode (Wood 1990).  
We should remark that this latter relation is based on  pulsation models
for variable AGB stars with oxygen-rich envelope compositions.

Due to the present lack of pulsation models suitably constructed
for C-stars, we have employed the  same period-mass-radius relation 
for TP-AGB models regardless of the actual C/O ratio.
As already mentioned, this fact implies 
that as soon as the third dredge-up makes C/O$\,>1$,
$P$ is predicted to increase because of the decrease in $T_{\rm eff}$
caused by the enhanced molecular opacities.

The physical soundness of this assumption could be verified only with 
the aid of pulsation models for C-rich compositions, using
adequate  $\kappa_{\rm var}$ opacities. Preliminary
calculations indicate that passing from C/O$\,<1$ to  C/O$\sim 2.5-3$,
the pulsation period actually almost doubles (Peter R. Wood, private
communication).
A detailed and quantitative investigation of this issue
is left to a future analysis.   
 
Another source of uncertainty is related to the incapability of the
our TP-AGB models, based on integrations of static envelopes, 
to account for the possible feedback between the third dredge-up and
HBB as shown by Stancliffe et al. (2004). In brief, deep dredge-up
episodes would actually increase $T_{\rm bce}$ compared to models
with shallower dredge-up, an effect which goes in the opposite
direction to that due to the  variable molecular opacities discussed in 
this work. In this context detailed AGB models with variable opacities
are needed to fully investigate the reciprocal feed-backs between the
third dredge-up and HBB.

\section{Summary and conclusions}
\label{sec_final}

This study has shown that
hot-bottom burning may be weakened, quenched or even  prevented 
as a consequence of the third dredge-up in intermediate-mass AGB stars.

The effect of the third dredge-up against HBB   
is to be ascribed  to the response of the outer layers 
(i.e. convective envelope and atmosphere) to the  
drastic change in the molecular opacities at the transition from 
C/O$ < 1$ to C/O$ > 1$. 
Essentially two factors contribute, namely:
\begin{itemize}
\item  The sudden decrease  of $T_{\rm eff}$ 
which accelerates the attainment of larger mass-loss rates, hence the 
reduction of the envelope mass;
\item The decrease of the temperature at the base of the convective envelope,
which weakens the nuclear reaction rates of the CNO-cycle.
\end{itemize}

These circumstances are expected to take place
{\sl only under specific conditions}, namely
i) in the early stages of the TP-AGB  phase,  
the third dredge-up is able to convert the AGB star into a carbon star 
before hot-bottom burning becomes strong enough to efficiently destroy 
the newly dredged-up carbon in favour of nitrogen, and ii)
the metallicity is high enough to allow molecules to form
in sufficient concentrations in the atmosphere.

Our calculations indicate that such  conditions  are likely  met 
in AGB models within a relatively small range of stellar masses, 
i.e. $M\approx 3.0-4.0 M_{\odot}$ and for metallicities $Z\ga 0.001$,  
 while they should not apply
to more massive models where any early transition to the C-rich domain
is prevented by HBB itself, and to more metal poor models for which
molecules do not play an important role. 

Despite the narrowness of the involved mass range, 
it has been shown that the new results should affect a sizable
fraction, say $\ga 30-50 \%$, of the whole group of massive AGB
stars that could undergo HBB.

The impact of such effect, so far neglected by full AGB models, 
may be large, so as to likely modify the current predictions 
(e.g. nucleosynthesis, mass range, lifetimes, chemical yields)
for the most massive carbon stars, that are 
predicted to populate the bright wing of the carbon star luminosity
function. 
These luminous carbon stars are expected to exhibit the fingerprints 
of the HBB nucleosynthesis and they could explain the existence of
particular groups of observed objects, such as the lithium-rich
carbon stars (Smith et al. 1995), and possibly those among the J-type carbon
stars that are found to be brighter than N-type stars (Morgan et
al. 2003). Moreover, the new results could affect the predicted
chemical abundances of planetary nebulae (PN) produced by
intermediate-mass progenitors, i.e. by reducing the mass range
of those that are usually associated to type I PNe 
(with measured over-abundances of He {\sl and} N).
 
Finally, our study indicates that
the impact of variable opacities on the yields of the 
lowest metallicity massive AGB models ($Z \le 0.00$1) 
is probably minor,  
due to the high dependence of molecular formation
on  temperature.
  
Therefore, in order to perform a reliable comparison between 
models and observations and to evaluate the role of these
stars in the chemical enrichment of the interstellar medium,
new  AGB evolutionary calculations, including variable molecular
opacities, are demanded.

\begin{acknowledgements}
P.M. thanks the anonymous referee and L\'eo Girardi for
their remarks that have contributed
to improve the original version of the paper.
This study was supported by the University of Padova (Progetto di
Ricerca di Ateneo CPDA052212).
 
\end{acknowledgements}

\end{document}